\documentstyle[referee,epsfig]{mn2e}
\newcommand{\degree}{{\rm o}}

\begin{document}

\title[Consistency of Bulk Flow Surveys]{Bulk Flows from Velocity Field Surveys: A Consistency Check}
\author[Sarkar, Feldman \& Watkins]{Devdeep Sarkar$^{\dagger,1}$, Hume A. Feldman$^{\dagger,2}$ \& Richard Watkins$^{\star,3}$\\
$^\dagger$Department of Physics and Astronomy, University of Kansas, Lawrence, KS 66045.\\
$^\star$Department of Physics, Willamette University, Salem, OR 97301\\
emails: $^1$dsarkar@uci.edu;\, $^2$feldman@ku.edu;\, $^3$rwatkins@willamette.edu}

\maketitle
\begin{abstract}
We present an analysis comparing the bulk--flow measurements for six recent peculiar velocity surveys, namely, ENEAR, SFI, RFGC, SBF and the Mark III singles and group catalogs. We study whether the direction of the bulk--flow estimates are consistent with each other and construct the full three dimensional bulk--flow vectors for each survey.  We show that although the surveys differ in their geometry, galaxy morphologies, distance measures and measurement errors, their bulk flow vectors are expected to be highly correlated and in fact show impressive agreement in all cases. We found a combined weighted mean bulk motion of $330$ km s$^{-1}$ $\pm 101$ km s$^{-1}$ toward $l= 234^{\degree}\pm 11^{\degree}$ and $b=12^{\degree}\pm 9^{\degree}$ in a sphere with an effective depth of $\sim4000$ km s$^{-1}$.
\end{abstract}


\section{Introduction}
\label{intro}

The analysis of peculiar velocity fields of galaxies and clusters is one of the most effective ways of probing mass fluctuations on $\sim$ 100 h$^{-1}$ Mpc scales (h being the Hubble constant in units of 100 km s$^{-1}$ Mpc$^{-1}$). Studies of peculiar velocities can be used to constrain the amplitude of  mass power spectrum on scales others than those probed by redshift surveys and those sampled by anisotropies in the CMB (e.g. Zaroubi et al.; 2001; Freudling et al. 1999). 
Originally motivated by the invention of distance indicators  based on intrinsic relations between galaxy observables \cite{FJ,TF} that are {\it independent of redshift}, the field has, in recent years, experienced great progress toward constructing large and homogeneous redshift-distance samples of galaxies and clusters. Although the analyses of early redshift-distance surveys of spiral galaxies \cite{aaronson} and of elliptical galaxies (e.g., Lynden-Bell at al. 1988) led to the development of several statistical methods for analyzing peculiar velocity data \cite{7s,kaiser1988,fw1994,sw1995,wf1995}, these studies were hampered by the fact that they were based on relatively small and shallow data sets. 

The present generation of redshift-distance surveys consist of larger and higher--quality data sets of   both spiral (da Costa et al. 1996; Giovanelli at al. 1997a, 1997b; Haynes et al. 1999a, 1999b; Karachentsev et al. 2000) and early-type galaxies (da Costa et al. 2000a, 2000b). These new samples pave the path toward a possible resolution of many discrepancies found in earlier samples; however, some quantitative disagreements persist. The earlier statistical comparisons of the peculiar velocity fields derived from D$_n$-$\sigma$ and Tully-Fisher (TF) distances found significant difference between them (e.g., Gorski et al. 1989; Tormen at al. 1993). 

Based on the work of Kaiser (1988), Feldman \& Watkins (1994) formulated a linear analysis to calculate the theoretical expectation for bulk flow in large-scale surveys as a function of the geometry of the survey, its clustering properties, and the assumed power spectrum; and applied it to a volume-limited complete sample of 119 Abell clusters (Lauer \& Postman 1994, hereafter LP) to show that the power spectra considered were inconsistent with the LP measurement of bulk flow at the 95$\%$-97$\%$ confidence level. The formalism was later applied \cite{wf1995} to calculate a measure of correlation between results obtained from the Riess, Press, \& Kirshner (1995a, 1995b, hereafter RPK) and the LP samples. They found that the apparent lack of agreement between the two measurements could be explained by the fact that both the LP and RPK samples were dominated by noise and incomplete cancellation of small scale motions.

More recently, in an analysis of the ENEAR sample (da Costa at al. 2000a, 2000b), the results obtained by Borgani et al. (2000) pointed toward a statistical concordance of the velocity fields traced by spiral and elliptical galaxies, with galaxy distances estimated using TF and D$_n$-$\sigma$ distance indicators, respectively. Following the method described in Feldman \& Watkins (1994) and Watkins \& Feldman (1995), Hudson et al. (2000) also showed that the bulk flows measured from four different surveys (SMAC, SC, LP10k and SNIa) were consistent with each other. 
Reconstruction techniques which compared the velocity to those predicted from galaxy redshift distributions (\cite{zaroubi2002,pike2005,pp2006}) show that we can get consistent results for $\beta=\Omega_m^{0.6}/b$ where $\Omega_m$ and $b$ are the mass density and linear bias parameters, respectively. These papers do not directly compare the surveys, but rather calculates best fit for $\beta$ given a velocity field. Pairwise velocity comparisons between various samples (\cite{pairwise2}) also show that different velocity samples produce consistent statistical results for $\Omega_m$ and $\sigma_8$ the standard deviation of density fluctuations on a scale of $8h^{-1}$Mpc. The fact that they give consistent results does not necessarily indicate consistency between the surveys, since the agreement is indirect and there is no attempt to quantify the consistency between the surveys and the field. Agreement of a specific statistical characteristic between data sets need not mandate consistent data sets.

In the present {\it paper}, we calculate the theoretically expected correlation between the estimates of the bulk flows of samples of galaxies in four recent surveys, namely, ENEAR (da Costa at al. 2000a, 2000b), SFI \cite{giovanelli1994,dacosta1996}, RFGC \cite{karachentsev2000}, SBF \cite{SBF} and the Mark III catalogs \cite{willick1997}.  We also introduce an analytical method to calculate the likelihood that two surveys both sample the same large scale flows, that is, we study whether measurement errors and differences in the distribution and morphology of galaxies in the surveys can statistically account for the differences in the directions of bulk--flow vectors.   Further, we  construct the 3-dimensional bulk flow vectors for all the surveys mentioned above and calculate the actual dot products of the estimates of the bulk flows obtained from these surveys in order to discuss their consistency. In $\S$ 2, we describe the theoretical background of velocity fields; we explain in detail the formulation of our analysis in $\S$ 3. A description of the surveys considered in our analysis is given in $\S$ 4; We then discuss our results in $\S$ 5 and conclude in $\S$ 6.   We argue, in the context of the EBW \cite{EBW} Standard cold--dark--matter (CDM) power spectrum, that our results also show a consistent statistical concordance.

\section{Physics of Velocity Fields}

In the context of the gravitational instability model of structure formation, the motions of galaxies are directly related to mass-density fluctuations. On scales of the surveys, the measured velocity of galaxies deviate from the Hubble expansion due to local mass distribution. Thus peculiar velocity surveys provide a unique method to probe the distribution of mass in the local universe. On scales that are small compared to the Hubble radius,  galaxy motions are manifest in deviations from the idealized isotropic cosmological expansion
\begin{equation}
cz = H_0r + \hat{{\bf r}} \cdot \left[{\bf v}({\bf r})-{\bf v}(0) \right]
\label{eq-cz}
\end{equation}      
where c is the speed of light, z is the redshift, $H_0$ is the Hubble constant, $r$ is the distance of a galaxy at the redshift z, $\hat{{\bf r}}$ is the the unit vector toward the galaxy, and ${\bf v}({\bf r})$ is the proper motion of the galaxy (at position ${\bf r}$) with respect to the comoving frame. This component of the overall motion of the galaxy is known as its {\it peculiar velocity}, arising from the gravitational attractions of surrounding overdensities. In Eq. (\ref{eq-cz}), ${\bf v}(0)$ is the peculiar velocity of the observer; It is standard practice to omit this term from the equation and to assume that redshift has been corrected to account for the motion of the observer.

The redshift--distance samples, obtained from peculiar velocity surveys, allow us to determine the radial (i.e., line--of--sight) component of the peculiar velocity of each galaxy:
\begin{equation}
v(r) = \hat{{\bf r}} \cdot {\bf v}({\bf r})= cz - H_0r
\end{equation}      

We assume that galaxies trace the large--scale linear velocity field ${\bf v}({\bf r})$ which is described by a Gaussian random field that is completely defined, in Fourier space, by its velocity power spectrum $P_v(k)$.
In the statistical model for peculiar velocities we define the Fourier Transform of the line-of-sight velocity $\hat{{\bf r}} \cdot {\bf v}({\bf r})$ such that:
\begin{equation}
\hat{\bf r} \cdot {\bf v}({\bf r}) = \frac{1}{(2\pi)^3} \int  d^3 {\bf k} \:  
	\hat{\bf r} \cdot \hat{\bf k} \:  v(\bf{k}) \: e^{i {\bf k} \cdot {\bf r}} 
\label{FT-vls}
\end{equation}

Due to the isotropy assumed in the Cosmological Principle, the statistical properties of $\hat{{\bf k}} v({\bf k})$ are independent of the direction of $\hat{{\bf k}}$, and so we may define the {\it velocity power spectrum} P$_v(k)$:
\begin{equation}
\left< v({\bf k})v^*({\bf k}^{\prime})\right>=(2\pi)^3P_v(k)\delta_D({\bf k}-{\bf k}^{\prime}),
\label{eq-Pv}
\end{equation}
where $\delta_D$ is a Dirac delta function, and the averaging on the left--hand--side is over directions of ${\bf k}$. 

In linear theory, the velocity power spectrum is related to the density power spectrum, P($k$), by
\begin{equation}
P_v(k)=\frac{H^2}{k^2} \: f^2(\Omega_{m,0},\Omega_{\Lambda}) \: P(k)\ .
\label{eq-Pv-P}
\end{equation}
$f(\Omega_{m,0},\Omega_{\Lambda})$ is the rate of growth of the perturbations at the present epoch and can be approximated as (e.g., Lahav et al. 1991):
\begin{equation}
f(\Omega_{m,0},\Omega_{\Lambda}) \approx \Omega_{m,0}^{0.6}
\end{equation}
where $\Omega_{m,0}$ is the cosmological density parameter for matter at the present epoch.

The power spectrum provides a complete statistical description of the linear peculiar velocity field. It should be noted that the above expressions are valid only on scales sufficiently large so that non--linearity can be neglected. In the present analysis, we consider the EBW parameterization of the linear CDM power spectrum \cite{EBW} 
\begin{equation}
        P(k) = \sigma_8^2 C
        k\Big(1+\big[6.4(k/\Gamma)+3(k/\Gamma)^{1.5}
        +(1.7k/\Gamma)^2\big]^{1.13}\Big)^{-2/1.13}
\end{equation}
where $\Gamma$ parameterizes the ``shape'' of the power spectrum and the overall normalization is determined by $\sigma_8$, the standard deviation of density fluctuations on a scale of $8h^{-1}$Mpc.  The constant $C$ is determined by the direct relation between $\sigma_8$ and the power spectrum.  For models where the total density parameter $\Omega=1$, the shape parameter is related to the density of matter, $\Gamma = \Omega_{m,0} h$. In the present analysis we use $\sigma_8 = 0.9$, $\Gamma = 0.21$ and $h=0.7$.

 \section{Modeling the Observational Data}

A catalog of peculiar velocities consists of a set of galaxies, labeled by an index n, for which we are given positions ${\bf r}_{n}$ and estimates of the line-of-sight peculiar velocities $S_n$ with  uncertainties $\sigma_n$.    For simplicity, we will make the assumption that observational errors are Gaussian distributed.   
Since linear theory only applies on scales comparable to the survey size, we focus our attention on the lowest order moments of a Taylor expansion of the velocity field ${\bf v}({\bf r})$.    Following Kaiser (1988), we model the velocity field as  a uniform streaming motion, or bulk flow,  denoted by ${\bf U}$, about which are random motions drawn from a Gaussian distribution with a 1--D velocity dispersion $\sigma_*$.   Although this model ignores the fact that small-scale motions, including those that are nonlinear, are correlated,   it is reasonable to assume that they effectively average out on the scales we are considering.    Since the value of $\sigma_*$ is not well determined by linear theory, we will treat it as a  parameter with a fixed value of 300 km  s$^{-1}$.    We have checked that our results are fairly insensitive to the exact value chosen for this parameter.   
Given these assumptions, the likelihood function for the bulk flow components is 
\begin{equation}
L(U_i) = \prod_n\frac{1}{\sqrt{\sigma_n^2+\sigma_*^2}} \: \exp \left( -\frac{1}{2} \: \frac{(S_n-\hat{r}_{n,i}U_i)^2}{\sigma_n^2+\sigma_*^2} \right)
\end{equation} 
where here and in subsequent equations repeated indices are summed over.
The maximum likelihood solution for the $i$th component of the bulk flow is given by  
\begin{equation}
U_i = A_{ij}^{-1} \: \sum_n \frac{\hat{r}_{n,j}S_n}{\sigma_n^2+\sigma_*^2},
\label{eq-Ui}
\end{equation} 
where 
\begin{equation}
A_{ij} = \sum_n \frac{\hat{r}_{n,i}\hat{r}_{n,j}}{\sigma_n^2+\sigma_*^2}
\label{eq-Aij}
\end{equation} 
Thus U$_i$ is the cross-correlation between the estimated line-of-sight velocity of the $n$-th galaxy and its position vector. For the catalogs considered, $A_{ij}$ is nearly diagonal, the off-diagonal terms being of order 10$\%$ of the diagonal ones.

In the model we are considering, the measured peculiar velocity of galaxy $n$ is related to the velocity field at the position of galaxy $n$ by
\begin{equation}
S_n = \hat{r}_{n,i}v_i({\bf r}_n)+\epsilon_n 
\end{equation} 
where $\epsilon_n$ is drawn from a Gaussian with zero mean and variance $\sigma_n^2+\sigma_*^2$. 
The fact that $\epsilon_n$ is statistically independent of the velocity allows the theoretical covariance matrix for the bulk flow components to be written as \cite{kaiser1988}
\begin{equation}
R_{ij}  = \left< U_i U_j \right> =  R_{ij}^{(v)} + R_{ij}^{(\epsilon)},
\label{eq-Rij}
\end{equation}
where the ``noise'' term can be shown to be \cite{kaiser1988}
\begin{equation}
R_{ij}^{(\epsilon)}  = A_{ij}^{-1}
\label{eq-Rije}
\end{equation}
and the ``theoretical" term can be written as the convolution of an angle-averaged tensor window function with the power spectrum 
\begin{equation}
R_{ij}^{(v)}  = 4\pi \int d k \: k^2 P_v(k) \: {\cal W}^2_{ij}(k)
\label{eq-Rijv}
\end{equation}
where 
 \begin{equation}
 {\cal W}^2_{ij} (k)= A^{-1}_{il}A^{-1}_{js} \sum_{n,m} {\hat r_{n,l} \hat r_{m,s}\over
(\sigma_n^2+\sigma_*^2)(\sigma_m^2+\sigma_*^2)}\int {d^2{\hat k}\over 4\pi}\ \left({\bf \hat r}_n\cdot {\bf \hat k}\ \ {\bf \hat r}_m\cdot {\bf \hat k}\right) \exp\left(i{\bf k}\cdot ({\bf r}_n- {\bf r}_m)\right)
\label{eq-WF}
\end{equation}

Our main goal in this {\it paper} is to figure out whether the surveys we consider are consistent with one another.   However, even if two surveys are measuring the same underlying velocity field, they will not necessarily give the same bulk flow.   This is both due to measurement errors in the peculiar velocities and the fact that each survey probes the velocity field in a different way.    This is most clearly seen by observing that each survey has different window functions (see below).    
In order to get an idea of how much correlation is expected between the estimates of the components of the bulk flows ${\bf  U}^A$ and ${\bf  U}^B$ of any pair of surveys (A,B) for a given power spectrum, we 
can calculate the correlation matrix $\left< {\bf U}^A{\bf U}^B\right>$ for the two surveys.   This is calculated in a similar manner to the covariance matrix, except that the two sums in the window function are now over two different surveys
\begin{eqnarray}
 {\cal W}^2_{ij} (k)= &(A^A)^{-1}_{il}(A^B)^{-1}_{js} &
 \sum_{n,m} {\hat r^A_{n,l} \hat r^B_{m,s}\over
((\sigma^A)_n^2+\sigma_*^2)((\sigma^B)_m^2+\sigma_*^2)} \nonumber\\ 
& & \times\int {d^2{\hat k}\over 4\pi}\ \left({\bf \hat r}^A_n\cdot {\bf \hat k}\ \ {\bf \hat r}^B_m\cdot {\bf \hat k}\right) \exp\left(i{\bf k}\cdot ({\bf r}^A_n- {\bf r}^B_m)\right)
\label{eq-WFcross}
\end{eqnarray}

The correlation matrix can then be used to calculate 
the normalized expectation value for the dot product of ${\bf  U}^A$ and ${\bf  U}^B$ \cite{wf1995}:
\begin{equation}
{\mathcal C} = \frac{\left< U_i^A U_i^B \right> }{\left( \left< U_l^A U_l^A \right> \left< U_m^B U_m^B \right> \right)^{1/2}}= \langle\cos\theta\rangle\ ,
\label{eq-C}
\end{equation}    
where $\theta$ is the angle between ${\bf  U}^A$ and ${\bf  U}^B$.   
$\mathcal C$ should be close to unity for highly correlated vectors, zero for vectors that are completely uncorrelated, and --1 if there is a high degree of anti-correlation.  It is important to realize that $\mathcal C$ carries information only about the correlation of the {\it directions} of the bulk flow vectors of the two surveys; however, it provides a convenient measure of how well the large scale velocity information contained in two surveys agree.

Given a value of ${\mathcal C}$ for two surveys (A,B) calculated using a given power spectrum, we can estimate the probability that the bulk flow vectors ${\bf  U}^A$ and ${\bf  U}^B$ will be separated by an angle greater than some $\theta_c$.   Our strategy is to think of the direction of  ${\bf  U}^A$ as scattering about the direction of ${\bf  U}^B$ in a two-dimensional space where $\theta$ is the radial distance.    Thus we can take $\theta$ to have a $\chi^2$ distribution with two degrees of freedom   
\begin{equation}
P(\theta)d\theta =  {\theta\over a^2} e^{-\theta^2/2a^2}\  d\theta.
\label{P-theta}
\end{equation}
The probability of measuring a value for $\theta$ greater than $\theta_c$ is then 
\begin{equation}
P(\theta>\theta_c) = \int_{\theta_c}^\infty P(\theta)\ \ d\theta.
\label{Pgt-theta}
\end{equation}
We can estimate the value of $a$ by using the fact that ${\mathcal C} = \langle \cos\theta\rangle \approx 1 - {1\over 2}\langle \theta^2\rangle$ for small $\theta$.     Since our $P(\theta)$ distribution has the property that $\langle \theta^2\rangle = 2a^2$, we can estimate 
\begin{equation}
a = \sqrt{1-{\mathcal C}}\ .
\label{width}
\end{equation}
This analysis ignores the small anisotropy in the covariance matrices for ${\bf  U}^A$ and ${\bf  U}^B$, but should be sufficient for our purposes.   

\section{The Surveys}

The formalism described above can be employed to test the consistency of all velocity field surveys. In this study, we have considered the following proper distance catalogs:\\

\newcounter{junk1}
\begin{list}{{\arabic{junk1})}}{\usecounter{junk1}\setlength{\rightmargin}{\leftmargin}}
\vspace{-0.1in}

\item  {\it Spiral Field I-Band (SFI)}: 
This is an all-sky survey (Giovanelli et al. 1994; da Costa et al. 1996; Giovanelli et al. 1998; Haynes at al. 1999a, 1999b), containing 1104 late-type spiral galaxies with I-Band TF distance estimates. It is a angular diameter limited survey and covers a volume out to $\sim$ 70 h$^{-1}$ Mpc. 

\item {\it Nearby Early-type Galaxy Survey (ENEAR)}:
This is an all-sky survey probing a volume out to 
$\sim$ 70 h$^{-1}$ Mpc. Although the survey contains data from different sources, it has been conducted by a single group and the data was analyzed by a single procedure and reached the same completeness level across the sky (da Costa at al. 2000a, 2000b). The sample contains 702 independent objects early--type elliptical galaxies and groups of galaxies brighter than m$_B$ $=$ 14.5 with D$_n$-$\sigma$ measured distances probing volume similar to the SFI survey. 

\item {\it Revised Flat Galaxy Catalog (RFGC)}: 
This catalog \cite{karachentsev2000} provides a list of radial velocities, HI line widths, TF distances, and peculiar velocities of 1327 spiral galaxies. This was compiled from observations of flat galaxies from FGC (Karachentsev, Karachentseva, \& Pernovsky 1993) performed with the 305 m telescope at Arecibo \cite{giovanelli1997} confined to the zone $0^{\degree}<\delta<+38^{\degree}$ accesible to the telescope.

\item {\it Surface Brightness Fluctuation (SBF) }:
This catalog \cite{SBF} employs I-band surface brightness fluctuation method and consists of 269 galaxies (both spiral and elliptical) reaching out to $\sim$ 4000 km s$^{-1}$, having a characteristic depth $\sim$ 12 h$^{-1}$ Mpc.

\item {\it Mark III catalog of singles:}\\ 
\noindent and
\item {\it Mark III catalog of groups:}
These catalogs \cite{willick1997} are a compilation of various disparate surveys that were recalibrated and compiled to provide some of the first reasonably dense and deep peculiar velocity surveys.  The Mark III Catalogs provide the observables for each object (i.e. redshift, magnitude, velocity width) and inferred distances derived from both the forward and inverse TF or $D_n-\sigma$ relations. Distances for both individual objects and groups are provided. The singles catalog has 2538 galaxies, while the group catalog has 1124 groups. The total survey depth is over 100 h$^{-1}$ Mpc with homogeneous sky coverage to $\sim 30$ h$^{-1}$ Mpc.
\end{list} 

The most important point to note here is that the above surveys are by and large independent of each other. They use different distance indicators, selection functions and survey geometries, and target different morphology of galaxies. Further, since our formalism weighs the galaxy contribution by the distance error, we do not expect distant galaxies to contribute much to our results and thus, the homogeneous Malmquist bias correction will not change our outcome, (see, e.g. \cite{pairwise1,pp2006}). 

As for the inhomogeneous Malmquist bias (IMB), since redshift distances are not affected, it will result in scattering the distances away from overdensities, leading to an appearance of an infall. This effect should not affect the bulk flow, though it may contribute to higher order moments. If we had found disagreement between the surveys, it would be very important to estimate the magnitude of the IMB since it may be the cause of the disagreement.  Since we found agreement, then it is likely that the IMB is smaller than our errors, since it was not sufficient to cause disagreement even though one would expect each survey to have a different bias. That said, we have conducted the following experiment: We have cut all galaxies above a certain radius
(for radii $r=40,50,60,70\ h^{-1}$ Mpc) and performed the analysis described above. Removing far away galaxies did not change the results significantly and the direction of the flow for these radii was within a standard deviation of the results quoted in this {\it paper}. These results are expected, since the galaxies are distance-error-weighted and thus, galaxies close to the edge of the surveys contribute relatively little to the overall flow.  We thus confirm our expectation that prominent features close to the edge of the surveys (e.g. Pieces-Perseus region) do not produce a spurious IMB bulk motion and systematically bias our results.

\section{Results}

We now present the estimates of the actual bulk flow vectors for all the different surveys. We use Eq. (\ref{eq-Ui}) to construct the Cartesian components of the full three-dimensional bulk flow vectors. Our results are tabulated in Table 1.
The uncertainties given for the bulk flow values in Table 1 are obtained from the noise part of the covariance matrix, $R^{(\epsilon)}$ (Eq. \ref{eq-Rije}).    Since $R^{(\epsilon)}$ is nearly diagonal for all of the surveys, we took the uncertainties in the $U_i$ to be approximately independent, so that the uncertainties are taken to be  $\sigma_{U_i} = \sqrt{R^{(\epsilon)}_{ii}}$.    These uncertainties are dominated by measurement errors in the individual velocities, although they also have small contributions from $\sigma_*$ and cosmic scatter.   


\begin{center}
\begin{tabular}{l|l|r|c|r|r|r}
\multicolumn{7}{c}{\bf Table 1} \\ \hline \hline
Survey 		& Method 		& N$\;\;$ 	& Effective Depth 	& U$_x\; \; \; $		& U$_y\; \; \;\; \; $	& U$_z\;\; \;\;$   	\\ 
       			&        		&   		& (km s$^{-1}$)    	& (km s$^{-1}$)    	& (km s$^{-1}$) 	& (km s$^{-1}$)    	\\ \hline \hline 
SFI			&	TF		&   1104	& $\sim$  4000		&     43$\pm$  37	&   -145$\pm$  35	&     57$\pm$  26  	\\ \hline
ENEAR		&$D_n-\sigma$&    702	& $\sim$  4000		&    154$\pm$  50	&   -246$\pm$  44	&     17$\pm$  39  	\\ \hline
RFGC		&	TF		&   1280	& $\sim$  6000		&    235$\pm$  38	&   -232$\pm$  39	&     31$\pm$  29  	\\ \hline
SBF			&	SBF		&    280	& $\sim$  2000		&    249$\pm$  58	&   -262$\pm$  46	&    163$\pm$ 29  	\\ \hline
Mark III singles&	Various	&   2538	& $\sim$  4500		&    202$\pm$  26	&   -230$\pm$  24	&     31$\pm$  22  	\\ \hline
Mark III groups	&	Various	&   1124	& $\sim$  4500		&    247$\pm$  34	&   -386$\pm$  31	&     79$\pm$  26  	\\ \hline
 \hline 
\end{tabular} 
\parbox{6in}{\small The Cartesian components of the full three--dimensional bulk flow vectors as well as the distance estimator method, the number of data points and the effective depth for each survey.}
\end{center} 

\vspace{3mm}

\begin{figure}
  \begin{center}
     \includegraphics[width=\linewidth]{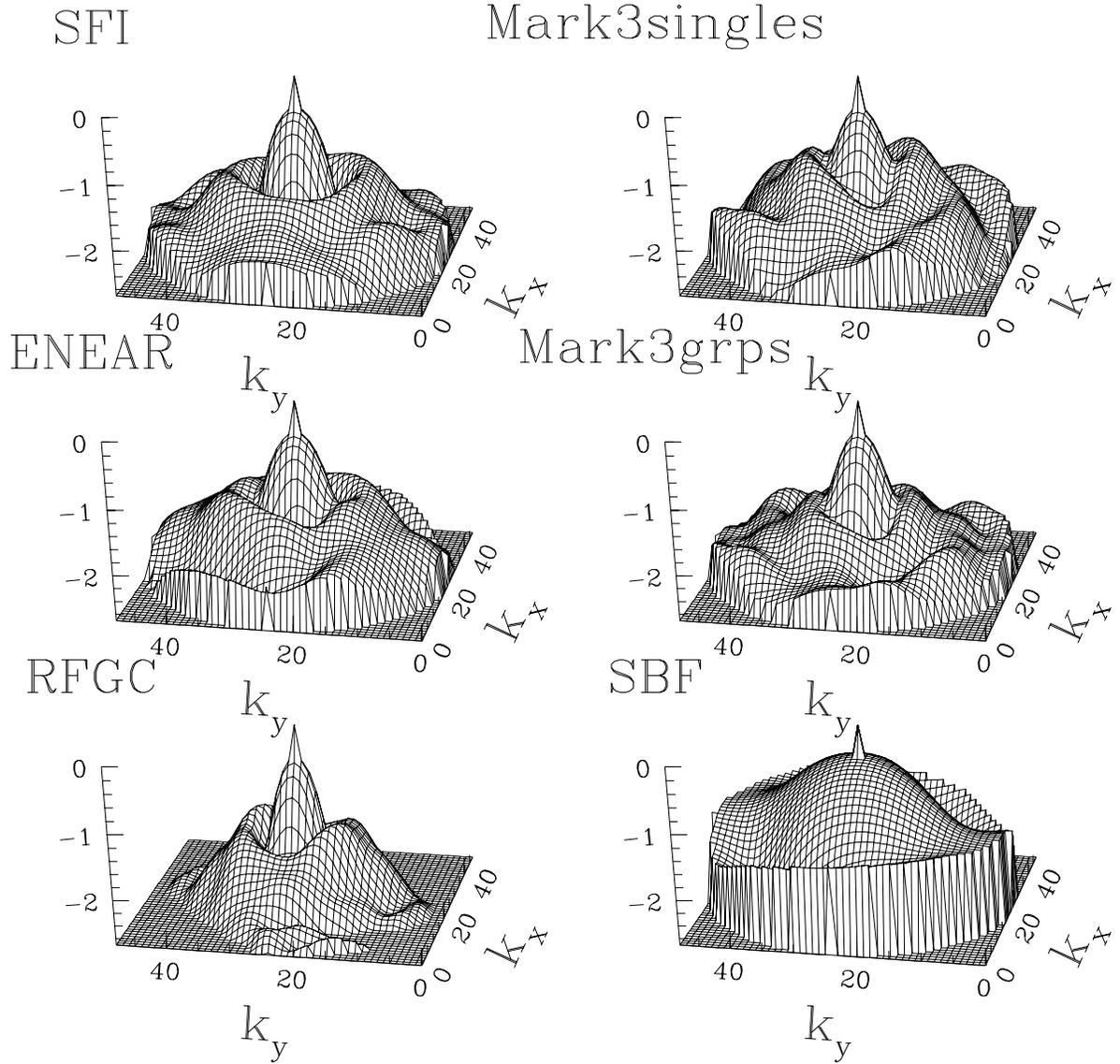}
        \caption{The logarithm of the trace of the normalized square tensor window functions in the $k_x-k_y$ plane from the six surveys we used in this study. As we can see, the central peak is similar in all surveys, suggesting that the large scale (small $k$) power is probed in a similar fashion. However, as we move away from the center, each survey samples the underlying power differently.}
        \label{fig-WF}
    \end{center}
\end{figure}

Each catalog surveys a different volume of space, it samples a small subset of the underlying population of galaxies, and uses independent spatial sampling techniques. Since the universe is homogeneous and isotropic on large scales and the sample volumes of the various data sets strongly overlap, these surveys react to the same underlying large-scale mass distribution. However,  the contribution from small scale nonlinearities differs from one catalog to another depending on the particulars of the survey. The window functions of these surveys, therefore, differ from one another, particularly on small scales. As a consequence, we don't expect the measured bulk flows from these surveys to be identical, even in the absence of peculiar velocity errors. Thus the bulk flow components listed in Table 1 are not strictly comparable.   However, if the samples are true representations of the flows of the underlying populations, they should be correlated.

For illustration purposes (Fig. \ref{fig-WF}) we show the tensor window functions (Eq. \ref{eq-WF}) for each of the surveys.    We chose to show the $ {\cal W}^2_{xx}$ in the $k_x,k_y$ plane (we have not performed the angle-averaging that is given in the equation).   The figure illustrates the fact that although all of the surveys have a similar central peak around $k=0$ which samples the large-scale power in a similar way, each   
survey samples the region of larger $k$ differently.    These differences in the window functions tend to decrease the correlation between the bulk flow vectors of the surveys.    

In Table 2, we show the value of $\mathcal C=\langle \cos\theta\rangle$ (Eq. \ref{eq-C}) for each pair of surveys (A,B), together with the inferred $a$ (Eq. \ref{width}) for the probability distribution of $\theta$.    We can see from these values that the directions of the bulk flow vectors for all of the surveys are highly correlated.  We also show the angle $\theta_c$ between the measured ${\bf  U}^A$ and ${\bf  U}^B$, and the probability of measuring an angle this large or larger, $P(\theta > \theta_c)$ (Eqs. \ref{P-theta}-\ref{Pgt-theta}).    These show that in general the results are consistent with one another for all pairs.    


\begin{center} 
\begin{tabular}{l|c|c|c|c|c|c|l} 
\multicolumn{6}{c}{\bf Table 2} \\ \hline \hline 
Survey 			& $\cos(\theta)$   & $\langle \cos\theta\rangle$ & $a$    &  $\theta_c$   &  $P(\theta>\theta_c)$\\
\hline\hline 
SFI--ENEAR         	&    0.92 &    0.91 &    0.30 &    0.40 &     0.42 \\ \hline
SFI--RFGC               &    0.85 &    0.90 &    0.31 &    0.55 &     0.21 \\ \hline
SFI--SBF                	&    0.91 &    0.78 &    0.47 &    0.44 &     0.64 \\ \hline
SFI--Mark III s         	&    0.88 &    0.89 &    0.33 &    0.49 &     0.33 \\ \hline
SFI--Mark III g         	&    0.95 &    0.92 &    0.27 &    0.33 &     0.49 \\ \hline
ENEAR--RFGC       	&    0.97 &    0.86 &    0.37 &    0.23 &     0.82 \\ \hline
ENEAR--SBF         	&    0.92 &    0.80 &    0.45 &    0.41 &     0.66 \\ \hline
ENEAR--Mark III s   	&    0.99 &    0.88 &    0.34 &    0.17 &     0.89 \\ \hline
ENEAR--Mark III g  	&    0.99 &    0.92 &    0.29 &    0.11 &     0.93 \\ \hline
RFGC--SBF		&    0.95 &    0.77 &    0.48 &    0.33 &     0.79 \\ \hline
RFGC--Mark III s	&    1.00 &    0.82 &    0.42 &    0.07 &     0.99 \\ \hline
RFGC--Mark III g	&    0.97 &    0.86 &    0.38 &    0.23 &     0.83 \\ \hline
SBF--Mark III s      	&    0.95 &    0.68 &    0.56 &    0.32 &     0.85 \\ \hline
SBF--Mark III g      	&    0.95 &    0.82 &    0.42 &    0.31 &     0.76 \\ \hline
Mark III s--Mark III g	&    0.99 &    0.92 &    0.28 &    0.17 &     0.83 \\ \hline
 \hline 
\end{tabular} 
\parbox{5in}{\small For each pair of surveys we show the value of the cosine of the angle ($\theta$) between their bulk flows, the expectation value of their dot product $\mathcal C$, the inferred width $a$ for the probability distribution of $\theta$, the critical angle $\theta_c$ and the probability of measuring an angle greater than $theta_c$.}
\end{center} 

To test our theoretical results and see in more detail the exact distribution of the bulk flow vectors, we conducted numerical experiments with the data. In one experiment we perturbed galaxies' positions, and hence also the peculiar velocities, using the reported errors in the distance measurements. Essentially we took each catalog and performed 1,000 Monte-Carlo realizations of the data using the measurement errors as the width of a Gaussian about the mean distance -- the proper distance reported. In another experiment we used the diagonal elements of the ``noise'' part of the covariance matrix as the variance of the individual components of the bulk flow vectors and did another 1,000 Monte-Carlo realizations where we drew values from a  Gaussian distribution, N($\mu$,$\Sigma^2$), with the individual components being taken as the mean $\mu$ and identifying $\Sigma^2$ as the variance obtained from $R^{(\epsilon)}_{ij}$ [refer to Eq. (\ref{eq-Rije})]. Both of our methods for calculating the spread in the components of the streaming solution yield similar results and agree with the errors reported in Table 1. 

We thus compared the bulk--flow vectors in two distinct ways, one was to measure the vectors directly from the data, monte--carlo the results and compare them, the other was to use the power spectrum to estimate the probability that two surveys measure different directions for the bulk flow. Since we used the power spectrum to calculate the likelihood that two surveys both sample the same large scale flows, one has to discuss the effect of cosmic scatter since that is part of the variance. However,  surveys that have good all-sky coverage, as our surveys do,   will have nearly identical contributions to their bulk flows from large scales;  the dominant part of the variance comes from small scale effects such as galaxy noise and distance measurement errors.    Thus only a very small part of the differences in direction of the bulk flows will come from large scale effects such as cosmic scatter. 
In order to remove the cosmic scatter part of the variance, we need to make a prior assumption, that we know the local velocity field from other sources, specifically reconstruction of density surveys. We chose not to use this prior and we show, in a model independent way, that velocity field surveys are consistent with each other and that they do sample the underlying large--scale in an unbiased, robust way.

\begin{figure}
  \begin{center}
     \includegraphics[width=\linewidth]{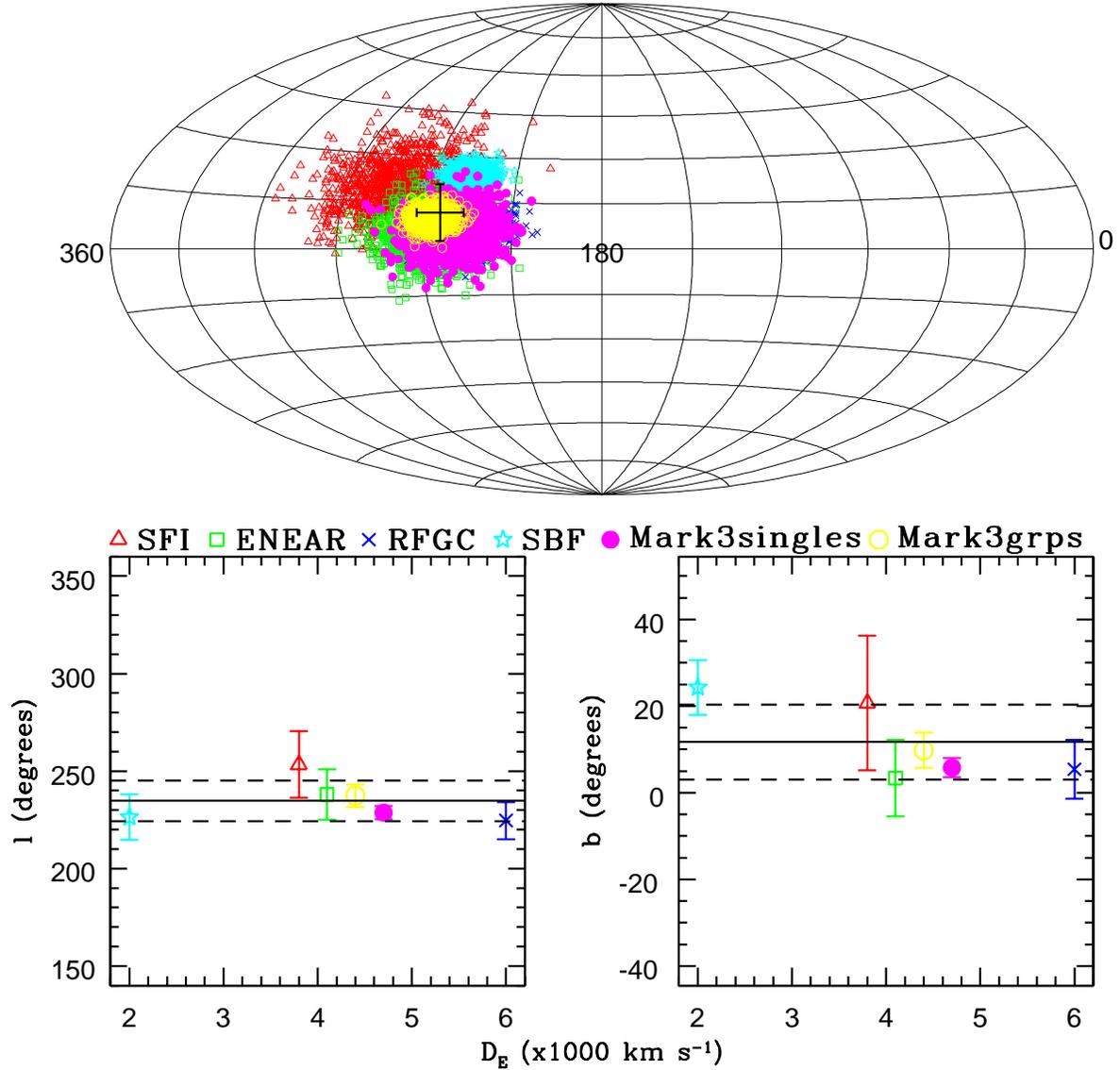}
        \caption{The top panel depicts the spread in the individual components of the bulk flow vector for all the six surveys, depicted in an Aitoff--Hammer Galactic projection, the cross indicates the weighted mean bulk flow direction. The bottom left and right panels show the galactic longitude and latittude, respectively as a function of the estimated effective depth ($D_E$) of the survey. The solid and dashed lines mark the mean and standard deviation, respectively, of the weighted mean results from the six catalogs of $l= 234^{\degree}\pm 11^{\degree}$ and $b=12^{\degree}\pm 9^{\degree}$}
        \label{fig-Ulb}
    \end{center}
\end{figure}

The bulk flow vector direction for each survey is given in Fig. \ref{fig-Ulb} where we plot the results from perturbing the distances, as described above, in the Aitoff--Hammer projection. It is clear from the figure that the bulk flow vectors for all surveys cluster about the same direction in the sky. Although the bulk flow components are not strictly comparable, it was hard to resist combining the results for all six catalogs to get an estimate of the mean bulk flow of  a sphere with an effective depth of $\sim4000$ km s$^{-1}$ to be approximately 330 km s$^{-1}$ $\pm$ 101 km s$^{-1}$ toward $l= 234^{\degree}\pm 11^{\degree}$ and $b=12^{\degree}\pm 9^{\degree}$ where $l$ and $b$ are the galactic longitude and latitude respectively. The value of the combined bulk flow vector was calculated by weighing the results by their errors, finding the weighted mean of the values for each survey. 
We would like to emphasize that this result should be taken with a grain of salt,  since the bulk flow is volume dependent and these surveys strongly overlap but do not strictly occupy the same volume.   We would also like to point out that the overall agreement between the bulk flow vectors of the different surveys may suggest that the internal sheer for the flows should be small, a conclusion we are testing in an upcoming paper (Watkins \& Feldman, 2006).

\section {Conclusion}

We have presented statistical analyses of the bulk flow measurement for six proper distance surveys. We have shown that the estimates of bulk flows obtained from these surveys are expected to have a high degree of correlation. Further, we have constructed the actual three dimensional estimates of the bulk flow vectors and shown that consistent results are obtainable from independent distance indicators, once they are applied to uniformly selected samples of galaxies. We find no statistically significant differences between the velocity fields mapped by different morphologies, galaxy types or distance indicators. 

We would like to stress that one should not be putting too much emphasis on comparing the components of the bulk flow vectors directly, since they are not really comparable.   This is especially true since the error bars reflect only the statistical and measurement errors and do not capture the differences in the bulk flows due to the fact that they probe the power spectrum differently.   Basically, we don't expect the bulk flow components to agree within the error bars shown in Table 1.  However, when we look at the {\it direction} of the bulk flow vectors, they do agree with each other remarkably well. Thus we conclude that all bulk flow measurements are consistent with each other given the errors as long as we allow for small scale alliasing and incomplete cancelations. A rough estimate of the (weighted mean) bulk flow from all surveys gives a flow of 330 km s$^{-1}$ $\pm$ 101 km s$^{-1}$ toward $l= 234^{\degree}\pm 15^{\degree}$ and $b=12^{\degree}\pm 9^{\degree}$.

This study clearly supports the notion that we have reached an era where velocity field data is consistent and robust across morphological types, selection criteria, survey geometry etc. Results from independent catalogs probe the same underlying large--scale power, though are subjected to different small--scale fluctuations. Unlike earlier, sparser surveys, the newer proper--distance surveys provide us 
with a dynamical probe of the large--scale structure which we can 
add to our growing arsenal  of data with confidence that the results reflect the cosmology we probe.

\noindent{\bf Acknowledgment:}
This research was supported by the University of Kansas General Research Fund (KUGRF). HAF has been supported in part by a grant from the Research Corporation.

\end{document}